# An experimental set up to probe the quantum transport through single atomic/molecular junction at room temperature

Biswajit Pabi* and Atindra Nath Pal*

Department of Condensed Matter Physics and Material Sciences,

S. N. Bose National Center for Basic Sciences,

Sector III, Block JD, Salt Lake, Kolkata - 700106 and

Email: biswajitpabi1@gmail.com, atin@bose.res.in

**Abstract:**

Understanding the transport characteristics at the atomic limit is the prerequisite for futuristic nano-electronic applications. Among various experimental procedures, mechanically controllable break junction (MCBJ) is one of the well adopted experimental technique to study and control the atomic or molecular scale devices. Here, we present the details of the development of a piezo controlled table top MCBJ set up, working at ambient condition, along with necessary data acquisition technique and analysis of the data. We performed conductance experiment on a macroscopic gold wire, which exhibits quantized conductance plateau upon pulling the wire with the piezo. Conductance peak up to ~ 20 $G_0$ ($G_0 = 2e^2/h$, e is the electronic charge and h being the plank's constant) could be resolved at room temperature. A well-known test bed molecule, 4, 4' bipyridine, was introduced between the gold electrodes and conductance histogram exhibits two distinctive conductance peaks, confirming the formation of single molecular junction, in line with the previous reports. This demonstrate that our custom-designed MCBJ set up is capable of measuring quantum transport of a single molecular junction at ambient condition.

## 1. Introduction

Despite many experimental challenges, electronic transport at the atomic scale have drawn significant attention in recent years. An atom or molecule suspended between two macroscopic metallic leads offers an atomic scale laboratory [1]. At this length scale, the electronic transport occurs primarily in the ballistic regime [2,3] in contrast to the diffusive transport observed in the bulk and the electronic transport is governed by quantum mechanics. A comprehensive and quantitative understanding of transport in these systems are addressed in a recent review by Michael Thoss et al. [4][5]. Many fascinating phenomena arising due to quantum mechanical effects may lead to modern technological implementation like quantized resistors, capacitors[6], switches[7], spin filter[8,9], molecular rectifier[10,11] and so on. Molecules being complex objects with multiple chemical configuration and vibrational degrees of freedom, understanding electronic transport through a single molecular junction is still an open topic of research [12–17]. However, Ferdinand Evers et al. summarizes the advances and challenges in single molecular transport by critically concerning theoretical and experimental findings so far [18]. Controlling the transport functionalities of the molecular junction via external gating[19,20] offers a promising strategy to develop high performance molecular devices such as molecular switches [21], molecular transistor[22,23] and thermoelectric devices[24], single electron logic calculator (SELC)[25].

One immediate question is how to wire a single molecule (with a typical size ~ 1 nm) to the electrodes. Due to the limited resolution in the existing micro fabrication techniques like electron or focused ion beam lithography, creating a gap size below ~ 10 nm is practically challenging. One way to circumvent this problem is to create a nanometer size gap through electromechanical methods where the gap with a possibility to fine tune the gap between the contact electrodes. Many such methods were tried in literature e.g., scanning tunneling microscopy break junctions (STMBJs)[13,26,27], crossed wire[28], mechanically controllable break junctions (MCBJs)[12], electromigration break junctions (E-BJs)[29,30], nanopores[31,32], liquid metal junctions [31,33,34] to form atomic size contacts. Mechanically controllable break junction [12,35,36] (MCBJ) technique is one of the widely used technique to study the electronic [37–39], mechanical [40,41], thermal[42,43] and spin-dependent [8,44–46] properties of a molecular junction. In case of MCBJ technique, a macroscopic metallic wire with a weak spot at the middle is placed on top of a flexible substrate. Three-point bending mechanisms of the MCBJ ensures an increase in strain in the wire during bending, which is concentrated at the notched spot, until the wire breaks. By relaxing the bending and incorporating fine control of the gap separation by means of a piezoelectric actuator, atomic junctions can be formed and broken repeatedly. Moreover, the piezo can be stopped at a desired configuration to carry out I-V characteristics. Small value of the "attenuation factor" ($\Delta = 6ut/L^2$; $t=$ thickness of the substrate, $u=$ length of the weak spot, $L =$ length between the counter support of the three point bending mechanism) defined as the ratio between gap separation and displacement of the piezo, makes it suitable to manipulate the conformation of the junction with a sub Angstrom resolution [47] and provides excellent mechanical stability. Moreover, controllable breaking and making cycles help to collect huge number of independent traces which can provide a good statistical description of the junction. By creating conductance histogram from these traces one can obtain the most possible conductance values for an atomic or a molecular junction. For a metallic junction, it was shown that the conductance histogram is the fingerprint of the available valence orbitals of the metal. Several metals have been characterized using MCBJ technique and statistically these exhibit different conductance value for single atomic contact, e.g.; Gold (Au) -1.0 $G_0$, Silver (Ag) - 1.0 $G_0$, Copper (Cu) - 1.0 $G_0$, Platinum (Pt) - 1.6 $G_0$, Aluminum (Al) ~ 0.8 $G_0$ (< 1.0 $G_0$) , Iron (Fe) - 2.1 $G_0$, Niobium (Nb ) - 2.3 $G_0$, Lead (Pb) - 1.7 $G_0$ ,Sodium (Na) - 1.0 $G_0$ and so on[48–52]. Wave-nature of the electron is considered to be responsible for conductance-quantization as the dimensions of the conductor is comparable to the de Broglie wavelength of the electrons at the Fermi surface and electrons can traverse across the conductor ballistically.

Here, we demonstrate an experimental set up based on MCBJ technique comprises of relatively low-cost materials instead of conventional electric motor or highly sensitive mechanical gear, reducing the fabrication cost significantly. Moreover, the preparation and assembly of the sample do not require any advanced techniques such as lithography or clean room facility. Our experimental set up is extremely stable and capable of probing quantum mechanical phenomena even at room temperature. Before entering into the experimental observations, a brief summary of quantum transport theory for an atomic conductor is presented. Initial characterization with gold atomic contact demonstrate the conductance quantization at integer multiple of $G_0$, in line with previous report [48]. Further measurements with 4, 4' bipyridine molecular junction reveal two conductance peak ($9.33 \pm 0.29 \times 10^{-4}$ $G_0$ and $4.28 \pm 0.07 \times 10^{-4} G_0$)[53–55] in the conductance histogram which in turn validate the binary conducting switching behavior of 4, 4' bipyridine molecular junction[38].

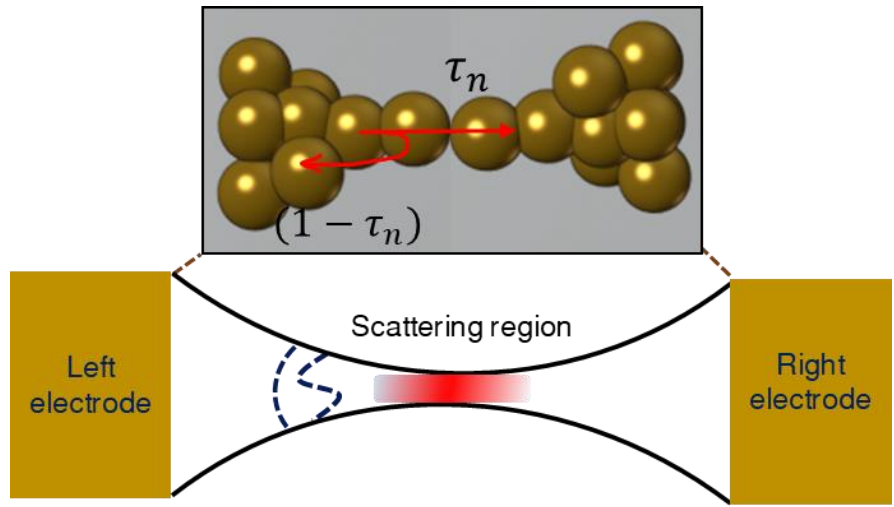

**Figure. 1. Landuer approach:** Landuer formalism of scattering approach to describe the electronic transport properties of atomic junction.Upper panel demonstrates the atomic junction (schematically) for which part of the wave is reflected and rest is transmitted, depending on the available conduction channel determined by the chemical nature of the atom.

## 2. Conductance in the quantum limit

Prior to the discussion of experimental part, it is convenient to discuss the transport mechanism of atomic scale junctions theoretically. Conductance through a few atoms to single atom or molecule is fundamentally different from its macroscopic counterpart and Ohm's law does not hold at this point. In case of atomic or molecular junction, characteristics length scale is smaller than the mean free path (the distance between two successive elastic collisions with static impurities) and the transport is assumed to be ballistic in nature. Momentum of the electron is considered to be constant in that regime and is only limited to the scattering from the boundaries of the contacts. Most celebrated theoretical formalism till date to describe the conductance of atomistic contact is the Landauer formalism of scattering approach, shown schematically in fig. 1[56]. As per this approach, a transport problem can always be considered as a scattering problem ignoring inelastic interactions. It essentially demonstrates that electrical conductance is intimately related to the transmission probability of the electron to cross the system and conductance can be expressed as,

$$G = \frac{2e^2}{h} \sum_{n=1}^{N} \tau_n = G_0 \sum_{n=1}^{N} \tau_n \qquad [1]$$

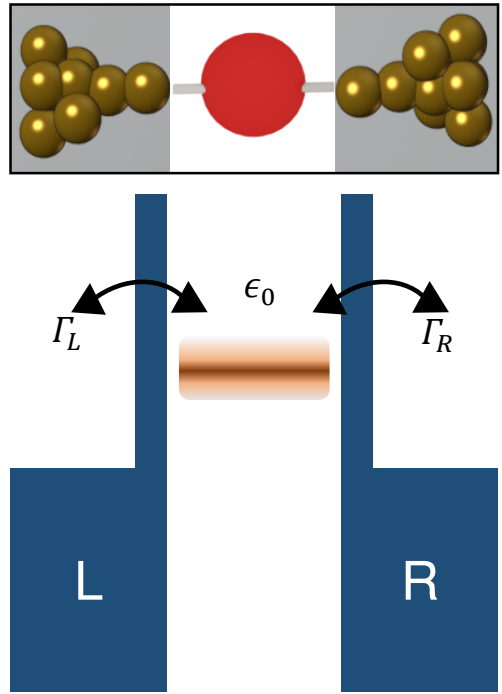

**Figure. 2. Energy level diagram:** Energy level diagram of a molecular junction, for a situation where transport is dominated by a single level, $\epsilon_0$. L and R denotes the left and right electrode. Upper panel shows a metal-molecule-metal junction schematically.

It is known as multi-channel generalization of Landauer formula[57] where conductance is viewed to be carried by N independent *conductance channels*. Each channel has a transmission probability $\tau_n$ and for a fully open channel has a conductance of $G_0$ ($G_0 = \frac{2e^2}{h} \equiv 12.9\ K\Omega$), known as *"quantum of conductance"*. This is an important contradiction compared to a macroscopic conductor where it is obvious to expect the zero resistance for a perfect conductor. Conductance channels of the atomic contacts are determined by the chemical nature of the atom and the number of active channels corresponds to the number of valence electrons[48] .

So far, our concern is limited to atomic contacts only. In this subsection, we will shed light on the transport mechanism through a molecular junction. Description of a molecular junction is rather complex compared to the atomic junction due to its multiple chemical conformations and vibrational degrees of freedom. Electronically a molecule is different from its gaseous phase, when placed in a junction. For simplicity we shall discuss the coherent transport model, a well-accepted model in the community to describe the current-voltage characteristics of a molecular junction, in which electrons flow elastically through the molecule without exchanging any energy. It is assumed that inelastic interaction takes place only well inside the electrodes, but not inside the molecule so that the phase information is conserved. In principle, different molecular orbitals can participate in the electron transport simultaneously. Some possible conduction mechanism includes direct tunnelling, Fowler-Nordheim tunnelling, thermionic emission or hopping transport [58,59]. In experiment, a particular mechanism is identified by studying the shape of current-voltage characteristics and their temperature dependence. For the sake of simplicity, we schematically represent the model using a single level resonant tunnelling model in fig. 2. Level position is denoted by $\epsilon_0$, the separation between the Fermi level and nearest molecular orbital in energy scale. Other key parameter in this model is the scattering rates $\Gamma_{L,R}$ which essentially demonstrates the strength of the coupling to the metallic leads $(L, R)$. I-V characteristics (calculated by considering the Landauer approach) adopts the following form,

$$I = \frac{2e}{h} \int_{-\infty}^{+\infty} dE\, T(E,V)\left[f\left(E - eV/2\right) - f\left(E + eV/2\right)\right] \qquad [2]$$

Where, factor 2 is coming from the spin symmetry, $f(E)$ is the Fermi function and $T(E,V)$ is the energy and bias voltage dependent transmission coefficient, given by the Breit-Wigner formula,

$$T(E,V) = \frac{4\Gamma_L \Gamma_R}{[E - \epsilon_0(V)]^2 + [\Gamma_L + \Gamma_R]^2} \qquad [3]$$

This simple model explains why, in a highly coupled molecular junctions, the conductance may be significant, even though in many cases the central energy of the molecular level is located far from the electrochemical potentials of the leads.

## 3. Design of experimental set up

To prepare MCBJ sample, a notch (weak spot) is created at the center of a macroscopic metallic Gold wire (diameter~0.1mm, Alfa Aesar, 99.998 %) using a surgical blade (Swan Morton) which is controlled by a Z positioner (Holmarc, India) as shown in fig. 3c. Metallic wire with the notch at the middle is then fixed on top of a flexible substrate (phosphor bronze of thickness ~1mm), covered by a kapton sheet for electrical insulation. An epoxy glue (Stycast 2850 FT with catalyst 9) is employed at both side of the notch to rigidly fix the wire (shown in the fig. 3b) with the substrate. The stycast is advised to be put very close to the junction, preferably within 100 μm for better junction stability. It is then mounted in a home-made three-point bending configuration (Schematic: fig.3a and Real set up: fig. 3d), where the flexible substrate is fixed at the two ends. Bending the substrate at the middle using a Z positioner leads to a strain in the wire, concentrated primarily at the weak spot. Upon further increase of the strain leads to the breaking of the wire, forming two freshly exposed atomistic electrodes. Then the junction can be repeatedly broken and formed using a piezo electric actuator (Piezomechanik Gmbh, Germany (PSt 150/2x3/7)), attached to the rod of the z-positioner. The details of the components used are given in *Table I*.

The circuit diagram shown in fig. 3e is used to measure the electrical conductance of single atomic or molecular junction. To record the conductance vs. electrode separation traces, a triangular waveform is applied to the piezo electric actuator which essentially translates to a small displacement of atomic electrodes and continued the successive break-make cycle. For conductance measurement, a dc bias from the output channel of a 16-bit DAQ card (PCI 6221, National Instruments) is applied and the current is measured by the input channel of the DAQ card after amplification using a current to voltage preamplifier (SP 983, electronics lab, University of Basel). Data acquisition is performed with the help of a custom build LABVIEW program. To increase dynamic range, keeping the amplifier gain constant, a series resistance $R_s$(50KΩ in our experiment) is connected [60]. Thus, when the conductance of the junction is higher, current is determined by the series resistance and in case of extremely low conducting junction (Resistance ~ GΩ), contribution of the series resistance is negligible and can be ignored. Conductance of the junction is thus measured by measuring the current and the voltage drop across the junction following the equations:

$$V_{Bias} = (V_{Rs} + V_{junction}); \ G_{Junction} = \frac{I}{V_{Junction}} = \frac{I}{(V_{Bias}-V_{Rs})} \quad [4]$$

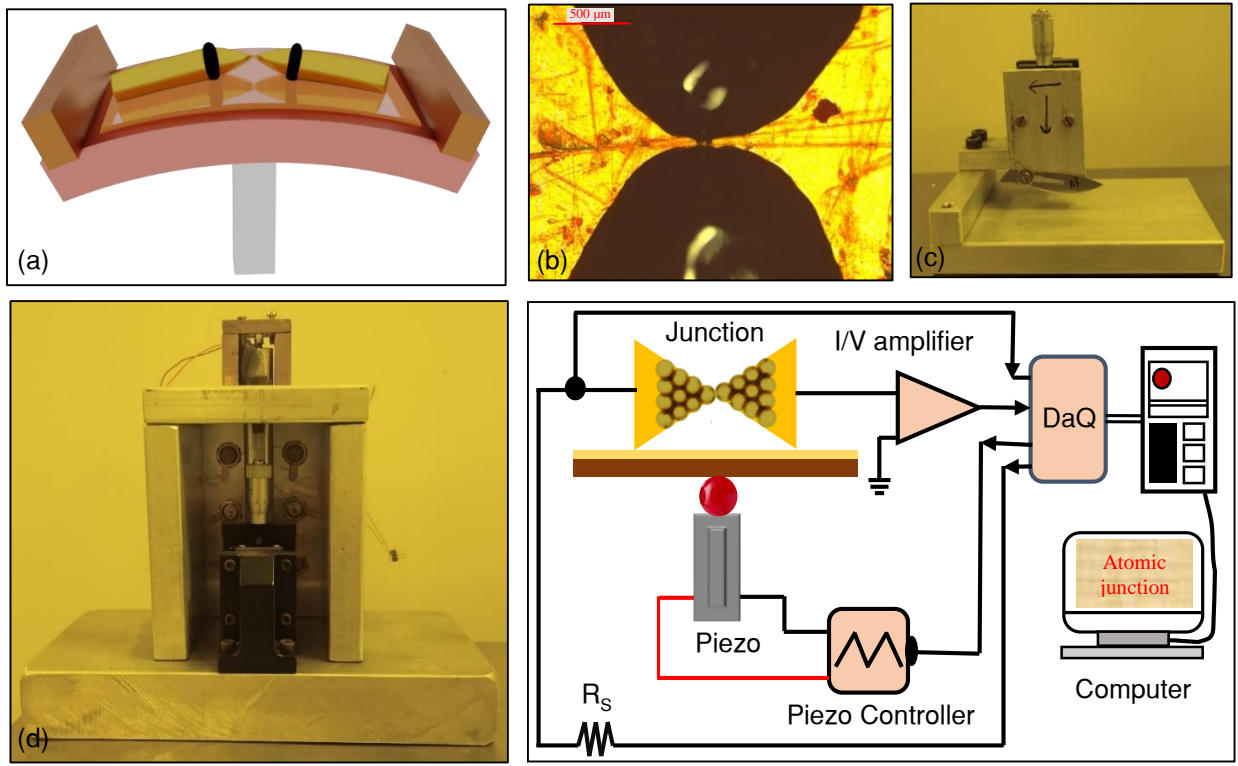

**Figure. 3. Experimental set up:** (a) Schematic representation of three point bending configuration of MCBJ technique. (b) A picture of the MCBJ sample with a notched gold wire where the two black spots are the epoxy (Stycast) that fixes the wire on top of the kapton foil. (c) The wire "notcher" designed to make the weak spot or notch in the wire by rolling the wire underneath the blade. (d) MCBJ set up fabricated to achieve the three point bending mechanism. (e) Schematic of electrical circuits, used for DC conductance measurements.

($V_{Bias}$ = Applied bias, $V_{Rs}$ = Voltage drop across the series resistance, $V_{Junction}$ = Voltage drop across the junction, $I$ = Current through the junction). By measuring current and the voltage drop across the junction one may obtain the conductance of the junction. An important consequence of this series resistance is that voltage drop across the junction is not constant throughout the measurements even though the applied bias is constant.

**Table I: Details of the components:**

| Materials or components | Company (Model) |
|---|---|
| Data acquisition System | National Instruments (PCI-6221 or PCI 4461) |
| Piezo stacks | Piezomechanik Gmbh, Germany (PSt 150/2x3/7) |
| Piezo controller | Piezomechanik Gmbh, Germany (SVR 150/1) |
| Current to Voltage preamplifier | Electronics lab, University of Basel (SP 983) |
| BNC cables and Resistance | element 14 |
| Z positioner | Holmarc Opto-Mechatronics ltd |
| Phosphor Bronze substrate | Local market |
| Kapton foil | Local Market |
| Stycast epoxy | Digi-Key electronics (2850 FT with catalyst 9) |
| Scalpel or surgical blade | Swann Morton (No.-24) |
| Aluminium Bar | Local Market |
| Electrode materials (Gold) | Alfa Aesar |
| Silver paint | SPI (05002-AB) |

4. **Data analysis tool**

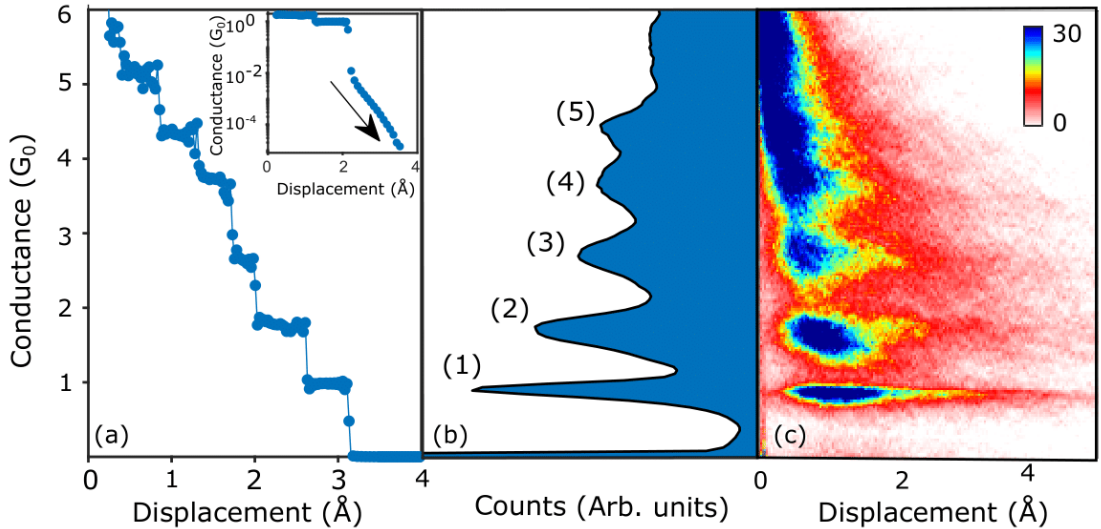

**Figure. 4. Atomic Junction:** (a) Characteristics breaking conductance trace of gold atomic junction. Inset shows the same traces in logarithmic scale to illustrate the quantum mechanical tunneling behavior indicated by a black arrow. (b) Conductance histogram constructed from 10000 consecutive breaking traces using 200 bins. Number in the histogram is used to assign those characteristics peak. (c) Two dimensional conductance-length histogram of same 10000 traces, constructed using 50 bins.

In this section, we will briefly explain the most frequently used data analysis tools: 1D conductance histogram and 2D conductance - displacement histogram. In case of break junction experiment, conductance of the junction is measured as a function of displacement between two electrodes, starting from bank of atoms to the vacuum tunnelling (i.e. no atom) by tiny and constant increment of gap size. Such a complete measurement cycle is called a *conductance trace* and thousands or even more such traces are collected to proceed further statistical analysis. Before entering next cycle, electrodes are always crashed up to several value of $G_0$ which essentially erase the memory from previous cycle. Statistical analysis of a large number of traces thus ensures the estimation of the most probable values of quantities like conductance, thermopower etc., from large number of independent events. 1D conductance histogram is constructed from the distribution of the conductance values of large number of traces. If a conductance plateau is frequently occurring in a certain conductance regime, it will be reflected as a peak in the histogram. By fitting those peaks in the histogram, most probable conductance values of the junction are obtained. In case of atomic junction, histogram is usually presented in linear scale in contrast to molecular junction where logarithmic binning is efficient since the single molecular conductance is lower than $1G_0$ and can be as low as $< 10^{-6}G_0$ [61]. Conductance histogram provides only the conductance values, ignoring information related to the displacement which is important to understand the evolution of the atomic or molecular junction during mechanical stretching or squeezing. Two dimensional conductance - displacement histogram is thus developed to look into the correlation between the conductance and displacement in a statistical manner. First, a conductance value is assigned for each trace as the origin of the distance axis i.e. zero distance point. Each point in the traces is then contributes to one of the 2D bins of the histogram, defined by conductance and distance from the starting point. The resulting histogram can then be considered as a stack of many traces, placed on top of each other and is helpful to figure out the most common feature during stretching. MATLAB scripts are used to carry out the statistical analysis.

## 5. Results and discussion

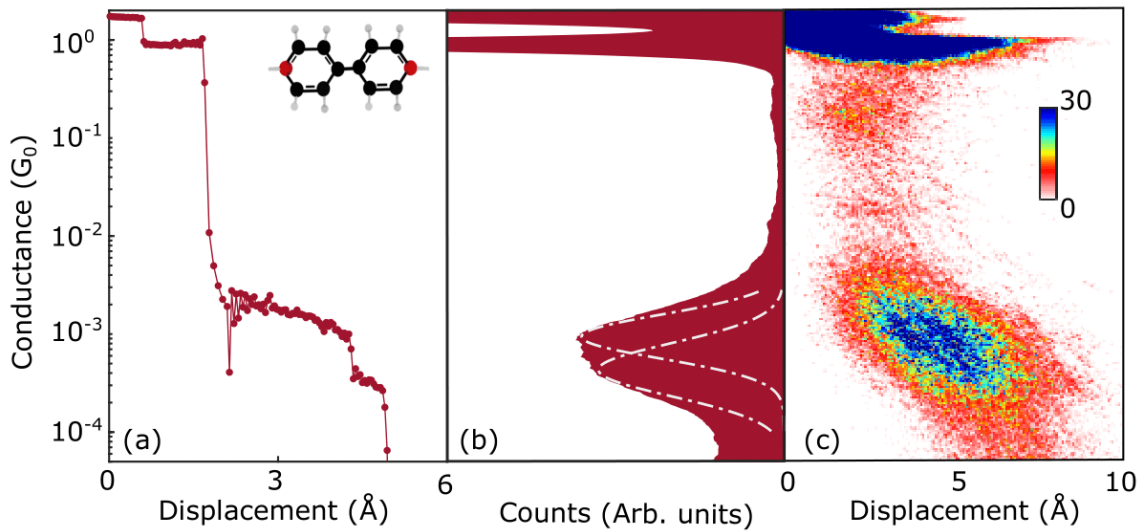

**Figure. 5. Molecular junction:** (a) conductance trace, (b) logarithmically binned conductance histogram and (c) two dimensional conductance-displacement histogram of 4, 4' bipyridine single molecular junction, connected to gold electrodes. 5000 consecutive trace is employed here to generate the 1D and 2D histogram using 100 and 50 bins per decade respectively. Gaussian fitting (white dash dot line) of the 1D conductance histogram yields the most probable conductance value of the molecular junction. Inset of (a) depicts the chemical structure of molecule considered.

A representative conductance traces of gold nano-junction at ambient condition, obtained using our home made MCBJ set up, is shown in fig. 4a. The curve reflects the evolution of particular atomic configuration upon stretching, during which conductance decreases in a series of sharp, vertically descending steps, known as *conductance plateaus*. As the number of atoms is reduced, the number of channels available for conduction also decreases and conductance drops in a discrete manner. Transport behavior at this atomic scale cannot be explained by the Ohm's law as stated earlier. As the Fermi wavelength of the electron is comparable to the dimension of the constriction, one needs to consider quantum mechanical confinement effect. Conductance of the latest atomic configuration is 1.0 $G_0$ (equivalent to ~12.9 k$\Omega$) and it has been argued that this configuration corresponds to the single atom at its narrowest cross section [48]. When the junction is pulled further, exponential quantum mechanical tunneling behavior between two atomic lead is clearly visible and shown as a black arrow in the inset of fig. 4a. Conductance histogram of gold nano junction constructed from 10,000 consecutive breaking traces having 200 number of bins, is shown in fig. 4b. Peaks at the conductance values ~ 0.96 $G_0$ (1), 1.75 $G_0$ (2), 2.75 $G_0$ (3), 3.70 $G_0$ (4), 4.46 $G_0$ (5) correspond to the most probable values of the conductance occurring from different configurations. Close inspection of the data reveals conductance peaks even up to conductance ~ 20$G_0$ (See Appendix B). The peak near 1.0 $G_0$ (~0.95 $G_0$) indicates the conductance of single atomic gold junction, characteristics signature of s-metal having single conduction channel[48]. Figure 4c shows the 2D conductance-distance histogram constructed from the same number of traces used in fig. 4b. It can be seen that the 1$G_0$ plateau can be stretched up to ~ 2 Å without much change in conductance. It is important to mention that one can observe the quantum transport behavior at such a high temperature (~ 300K). Because of the inelastic electron-phonon scattering, one would expect that there will mixing of different quantum channels. However, recent shot noise measurement on atomic point contact of gold at 300K[62], a clear suppression of shot noise observed near G = 1.0 $G_0$, indicating the ballistic transport even at room temperature. It indicates that the inelastic scattering length is much larger than the dimension of the atomic point contact (~few Å) at room temperature and at a finite bias of few tens of millivolt range.

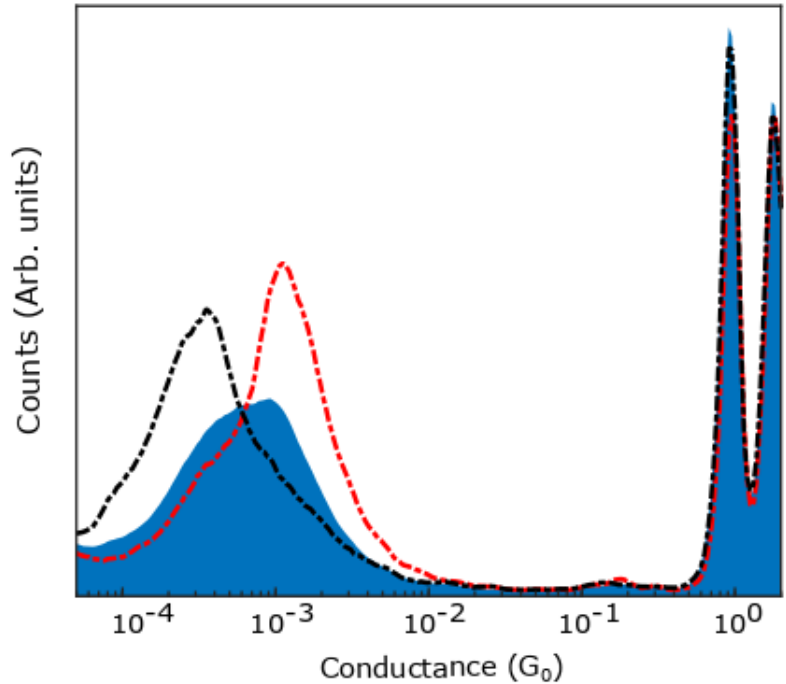

**Figure. 6. Conditional histogram:** 1D logarithmic conductance histogram of 4, 4' bipyridine single molecular junction where blue area plot is constructed from the traces having molecule. Red dash line & black dash line correspond to the histogram for which molecular plateaus lie in the high conductance regime & low conductance regime respectively.

After characterizing the set up with gold atomic junction, we will now discuss the possibility of studying the conductance of a molecular junction between two gold electrodes. Here, our choice is 4, 4' bipyridine (Chemical structure: Inset of fig. 5a), one of the well-studied molecules in the community [38,63–67]. For creating molecular junction, the molecule is evaporated on top of the notched metallic wire. Fig. 5a shows the typical breaking traces of 4, 4' bipyridine molecular junction in logarithmic scale where molecular plateaus (from $3 \times 10^{-3}$ $G_0$ to $2 \times 10^{-4}$ $G_0$) is observed along with the atomic plateaus at 1.0 $G_0$. To get a meaningful estimate of the conductance value of the molecular junction, 5000 such independent traces are analyzed and presented as a logarithmically binned normalized 1D histogram in fig. 5b (100 number of bins per decade). Histogram demonstrates a peak at 1$G_0$ which corresponds to Au atomic contact, associated with a prominent molecular characteristics ($5 \times 10^{-3}$ $G_0$ to $1 \times 10^{-4}$ $G_0$). The molecular conductance peak can be fitted with two Gaussian peaks ($9.33 \pm 0.29 \times 10^{-4}$ $G_0$ and $4.28 \pm 0.07 \times 10^{-4}$ $G_0$) (shown as white dash-dot line in fig. 5b). Conductance values of these two peaks are in good agreement with previous report [38]. To look into the conductance behavior during stretching, 2D conductance displacement histogram is generated (employing same traces as used in 1D histogram) by aligning them at 2.0 $G_0$ and using 50 bins per decade. In the 2D histogram, we observed that the molecular junction can be stretched up to ~ 8Å and the conductance decreases with stretching. The low conductance peak mainly originates from the configuration when the molecule is completely stretched[38]. To gain detail insight about the formation of these two different configurations, conditional histogram is considered and shown in fig. 6. Blue area plot demonstrates the histogram containing all the molecular traces. However, the line plots represent the histograms which are generated by selecting traces with only high conductance [high-G] peak (Red) or low conductance [low-G] peak (Black). It is evident that by selecting one of the peaks does not remove completely the other peak, rather we observe a small suppression. It can be shown that these two configurations are plateau length-wise anti-correlated, as shown in Ref [68]. Thus, statistical analysis of the experimental data confirms the successful formation of 4, 4' bipyridine molecular junction using our home built experimental set up.

## 6. Conclusions

In this work, we have demonstrated a simple and robust experimental arrangement based on MCBJ technique without using sophisticated electric motor or high-resolution gear to look into electronic transport behavior at the atomic scale. Initial characterization of our experimental set up with gold atomic junction shows well defined conductance plateaus at the integer multiple of $G_0$ ($G_0 = 2e^2/h$) which has been further verified by the statistical analysis of a large number of traces. We have also studied the 4, 4' molecular junction and happy to conclude that most probable conductance value of this molecular junction is in excellent agreement with the previous report. Thus, our experimental set up is good enough to probe the quantum transport behavior through atomic and molecular junction. Fabrication of this table-top set up will be more helpful for motivated high school students or lab course at the undergraduate or postgraduate level. Moreover, direct observation of quantum mechanical phenomena at room temperature and ambient condition will be a fascinating demonstration of quantum mechanics at our daily lives.

## 7. Acknowledgement

B. Pabi acknowledges support from DST-Inspire fellowship (Inspire code- IF170934) and A. N. Pal acknowledges the funding from Department of Science and Technology (grant no: CRG/2020/004208). We are grateful to Amit Ghosh for necessary machining using workshop at SNBNCBS. We acknowledge Ayelet Vilan for helping with the MATLAB code and Oren Tal for fruitful discussions.

**Appendix A: Calibration between the displacement (or electrode separation) and the voltage applied to piezo element.**

Experimentally, distance between two electrodes is manipulated by applying a voltage to the piezo electric actuator. Adopting the procedure from Ref. [69], exponential dependence of the current on the vacuum gap is used to make a rough calibration of the gap size. Linear relation between the piezo expansion that mean the voltage at piezo, $V_{piezo}$ and gap size, $\Delta$ (i.e. electrode separation or displacement) helps to define a simple calibration constant ($c$),

$$c = \frac{\Delta}{V_{piezo}} \quad (A1)$$

Tunnel current (I) between two electrodes which are separated by a distance $\Delta$ (provided applied voltage $V_0$ is smaller than the work function of the electrodes) can be expressed as[70],

$$I(V_0) = kV_0 e^{-2\Delta\sqrt{2m\phi/\hbar^2}} \quad (A2)$$

Where m is mass of the electron, $\phi$ is the work function of the electrode, $k$ is a constant related to the area of the electrode and to the electron density of states at the Fermi level and $\hbar$ is the reduced Planck constant. Resistance of the tunnel junction is thus,

$$R = R_0 e^{2\Delta\sqrt{2m\phi/\hbar^2}} \quad (A3)$$

Combining equation (A1) and equation (A3), we can write,

$$R = R_0 e^{2cV_{piezo}\sqrt{2m\phi/\hbar^2}} \quad (A4)$$

Slope m, of the logarithmic resistance with respect to piezo voltage is thus,

$$m = \frac{\partial(\ln R)}{\partial V_{piezo}} = \frac{\partial(2cV_{piezo}\sqrt{2m\phi/\hbar^2})}{\partial V_{piezo}} = \frac{\sqrt{2m\phi}}{\hbar} 2c \quad (A5)$$

$$\text{Calibration constant, } c = \frac{m*\hbar}{2\sqrt{2m\phi}} \quad (A6)$$

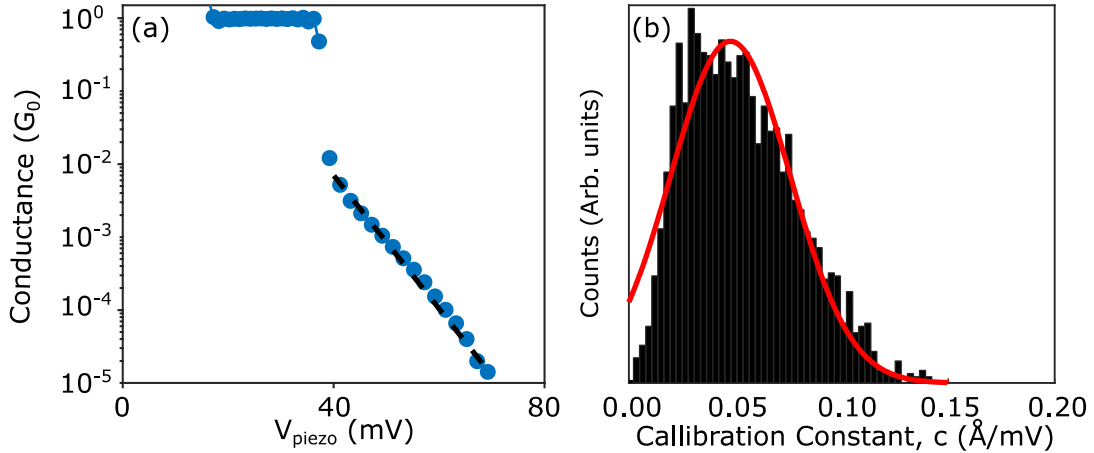

**Figure. 7. Calibration:** (a) Conductance traces of gold atomic junction where conductance is presented in terms of piezo voltage ($V_{piezo}$). Region is used for calculation of calibration constant is marked by the black dash line. (b) Histogram of calculated calibration constant, obtained from the 10000 traces of clean gold junction. Red line plot is the Gaussian fitting which elicits the most possible calibration constant ~ 0.053 Å/mV.

This expression is indeed very simple and clean electrodes follows an exponential behavior (shown in fig. 7a) of the current as a function of $V_{piezo}$, which would make this a suitable method to calibrate of the gap size with voltage applied to the piezoelectric actuator. A histogram of calibration constant (shown in fig. 7b) is generated using the calculated slopes of bulk number of make (or push) traces. Then, most frequent $c$ value is obtained by Gaussian fitting to this histogram and is used further to calibrate. However, major source of inaccuracy in this method is coming from the value of Work function used which is very much sensitive to the fine structure details of the junction[71] and also on the local environment[72].

**Appendix B: Gold atomic junction at higher conducting configuration.**

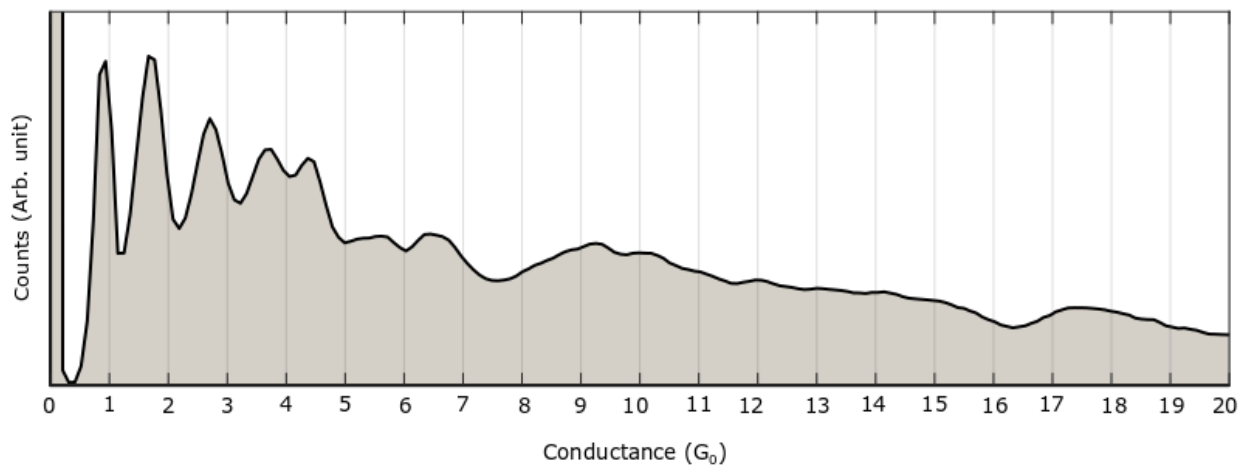

**Figure. 8. Atomic junction:** One dimensional conductance histogram of gold atomic junction is shown till conductance value upto 20$G_0$, constructed from the same 10,000 traces used in figure 4, using 200 bins. High conductance peaks can be resolved even upto ~20$G_0$.